\documentclass[aps,prd,twocolumn]{revtex4-2}

\usepackage{amsmath}
\usepackage{graphicx}
\newcommand{\bx}{{\hbox{\boldmath $x$}}}
\newcommand{\bu}{{\hbox{\boldmath $u$}}}
\newcommand{\bh}{{\hbox{\boldmath $h$}}}
\newcommand{\bq}{{\hbox{\boldmath $q$}}}

\newcommand{\sbn}{{\hbox{\scriptsize\boldmath $n$}}}
\newcommand{\tbn}{{\hbox{\tiny\boldmath $n$}}}
\newcommand{\bJ}{{\hbox{\boldmath $J$}}}

\begin{document}

\title{Extending Hamiltonian Formulation of Particle Motion in Perturbed Kerr Spacetime to various time parameterizations}
\author{Takafumi Kakehi$^{1,3}$}
 \email{takafumi.kakehi@yukawa.kyoto-u.ac.jp}
\author{Takahiro Tanaka$^{2,3}$}
\email{t.tanaka@tap.scphys.kyoto-u.ac.jp}
\affiliation{%
$^1$Yukawa Institute for Theoretical Physics$,$ Kyoto University  
}%
\affiliation{$^2$Department of Physics$,$ Kyoto University$,$ Kyoto 606-8502$,$ Japan}
\affiliation{$^3$Center for Gravitational Physics and Quantum Information$,$ Yukawa Institute for Theoretical Physics$,$ Kyoto University$,$ Kyoto 606-8502$,$ Japan}
\begin{abstract}
The Hamiltonian formulation with action-angle variables is very useful when considering the motion of particles undergoing a self-force reaction due to gravitational wave emission. Using the proper time as a parameter along the trajectory is considered to be appropriate when developing formal arguments, but these arguments can easily be extended to arguments with other time variables. The aim of this paper is to clarify the relations between the schemes with various parameterizations of the orbit.
\end{abstract}
\maketitle

\section{Introduction}
The two-body problem in general relativity is one of the key issues in modern physics. Binary systems are particularly important as sources of gravitational waves, providing critical information across a wide range of fields, from fundamental physics to astronomy. Among them, extreme mass ratio inspirals (EMRIs) are the key target for next-generation gravitational wave observatories like LISA\cite{amaroseoane2017laser}, Taiji\cite{Hu:2017mde}, TianQin\cite{TianQin:2015yph} or DECIGO\cite{Kawamura:2011zz}. 
Theoretical methods to predict the time evolution of EMRIs have been actively pursued since the discovery of the fundamental self-force formula\cite{Mino:1996nk,Quinn:1996am}, and many recent discussions focus on the use of action-angle variables~
\cite{Schmidt:2002qk,Hinderer:2008dm,Fujita:2016igj,Barack:2018yvs, Pound:2021qin, 
Blanco:2022mgd,Kerachian:2023oiw,Grant:2024ivt}.


Using the action-angle variables, 
the balance formulae that describe the change rates 
of the action variables were derived in \cite{Isoyama:2018sib}. 
On the other hand, it is convenient to use the
Mino time\cite{Mino:2003yg} in the actual computation, 
and the phase variables that uniformly lapse 
in Mino time are used to describe geodesics. 
Using these variables, a concise formula to evaluate 
the change rate of the Carter constant\cite{PhysRev.174.1559}, 
which can be related to the change rates of the 
action variables, was actually derived 
earlier in \cite{Sago:2005fn} than that based on the Hamiltonian formulation mentioned above. These two expressions 
look quite similar but they are not the same. 
The main purpose of this short paper is to explore the relation between these two expressions 
a little deeply. 

Also, it seems that there is a misunderstanding that 
the Hamiltonian dynamics and the action-angle variables are not defined in the formulation that uses Mino time as the time coordinate. We wish to correct this misunderstanding. Based on this notion, the formulae obtained in \cite{Sago:2005fn} can be directly derived, following the same simple logic used in \cite{Isoyama:2018sib}. 

This paper is organized as follows. In Sec. II we present a framework to discuss different choices of the time parametrization along the orbit in the Hamiltonian formulation. Then, the canonical transformation between action-angle variables in various schemes is examined in Sec. III. We mainly discuss the relation between two cases with the proper time and Mino time parameterizations, but the results apply to more general cases. 
In Sec. IV we comment on an alternative proof of the correspondence between the radiation reaction formulae 
in the context of the first-order self-force, which is based on the invariance of the asymptotic waveform 
between different schemes. This proof might not be explicitly mentioned before, but it was already recognized within the community as an explanation of why different calculation methods yield the same answer. 
In Sec. V we consider the case in which we use the external time as the parameterization of the trajectory. 
Section VI summarized this short paper.  

\section{Hamiltonian in arbitrary time parametrization}
We start with the re-parameterization invariant action of a point particle in an arbitrary given metric $g_{\mu\nu}$,
\begin{align}
  S=-\int ds\sqrt{-g_{\mu\nu}\frac{dx^\mu}{ds}\frac{dx^\nu}{ds}}\,.
\end{align}
Here, the particle mass is set to unity, for simplicity.
The conjugate momenta are defined by 
\begin{align}
 u_\mu&:= \frac{\partial L}{\partial (dx^\mu/ds)}\cr
   & =g_{\mu\nu}\left.\frac{dx^\nu}{ds}\right/
  \sqrt{-g_{\rho\sigma}\frac{dx^\rho}{ds}\frac{dx^\sigma}{ds}}\,.
\end{align}
An equivalent action which does not contain the square root is obtained just by moving to 
the Hamiltonian form as
\begin{align}
  S=\int ds\, \left[u_\mu \frac{dx^\mu}{ds}-H\right]\,,
\end{align}
with
\begin{align}
 H:=\frac{e}{2}\left(-\mu^2 +1\right)\,,
\end{align}
where $e$ is a Lagrange multiplier and 
\begin{align}
  \mu^2:=-g^{\mu\nu}u_\mu u_\nu\,. 
\end{align}
The Hamiltonian merely describes the constraint, 
$\mu^2=1$,
which means that $u_\mu$ is the ordinary four velocity in any choice of 
$e$. 
The consistency of the time evolution of the constraint does not give any condition on the choice of $e$. 
Therefore, $e$ remains arbitrary, unless we do not specify a gauge fixing condition. 
In this section, we will not impose any gauge fixing condition, and $e$ is specified by hand after taking the variations 
with respect to the dynamical variables. 

When we consider a motion in $n$ dimensions, and if we have $n$-commuting constants of motion, the $2n$-dimensional phase space is foliated by $n$-dimensional tori, along which there are $n$ non-trivial cycles, $C_a$\cite{arnold1989mathematical}.   
Then, the action variables are defined by the integral along respective cycles as 
\begin{align}
 J_a:=\frac1{2\pi}\oint_{C_a} u_\mu dx^\mu\,.
\end{align}
The definition of $J_a$ is also independent of the choice of $e$, as long as the 
constraint is satisfied. 
Here, we should notice that the constraint equation can be written in terms of $J_a$
as 
\begin{align}
 \mu^2(\bJ)=1\,,  
 \label{eq:constraint}
\end{align}
which defines the constraint surface ${\cal S}$. 
Only on ${\cal S}$ all descriptions with different choices of $e$ are equivalent. When we assume $e$ is just a function of time parameter $s$ given by hand, we can consider the motion apart from ${\cal S}$. 
In this case, $n$ constants of motion and hence the $n$-dimensional tori defined by the constants of motion depends on the choice of $e$. Therefore, the definition of $J_a$ also varies. We denote the angle variables conjugate to $J_a$ by $q^a$. 


From now on, we consider a perturbed spacetime with the metric defined by $g_{\mu\nu}+h_{\mu\nu}$.
Using the action-angle variables $(\bJ, \bq)$, which are related to the original variables $(\bu, \bx)$ via the generating function for the background metric $g_{\mu\nu}$, 
we derive the equation governing the time evolution of the action variables. 
The evolution equation of $\bJ$ is obtained as
\begin{align}
 \left\langle \frac{dJ_a}{ds}\right\rangle
 &=-\left\langle \frac{\partial H}{\partial q^a}\right\rangle\cr
 & =-\left\langle \frac{\partial H_{\rm int}}{\partial q^a}\right\rangle
 =- \frac{\partial \left\langle H_{\rm int}\right\rangle}{\partial q^a_{\rm ini}}\,,
\end{align}
where
\begin{align}
 H_{\rm int}:= - \frac{e}{2} h^{\mu\nu} u_\mu u_\nu\,, 
\end{align}
and $q^a_{\rm ini}$ is the initial value of $q^a$\cite{Isoyama:2018sib}. 
$\langle\cdots \rangle$ represents the long time average. 
We find that the flux formula can be obtained 
in any choice of the parametrization, although
the angle variables are different for a 
different choice of $e$, as we shall see more explicitly in the succeeding section. 

\section{Canonical transformation}
In the case of Kerr geodesic motion, 
we are interested in two particular cases: 
one is the case with $e=1$ and the other is 
$e=\Sigma:=r^2+a^2\cos^2\theta$. We put tilde 
to the variables in the latter case. 
For the parametrization of the orbit, we use 
$\tau$ and $\lambda$ for the respective cases. 
$\lambda$ is the so-called Mino time\cite{Mino:2003yg}. 
Using Mino time one can solve the radial and polar-angle motions independently. 

\begin{figure}
    \centering
    \includegraphics[width=\linewidth]{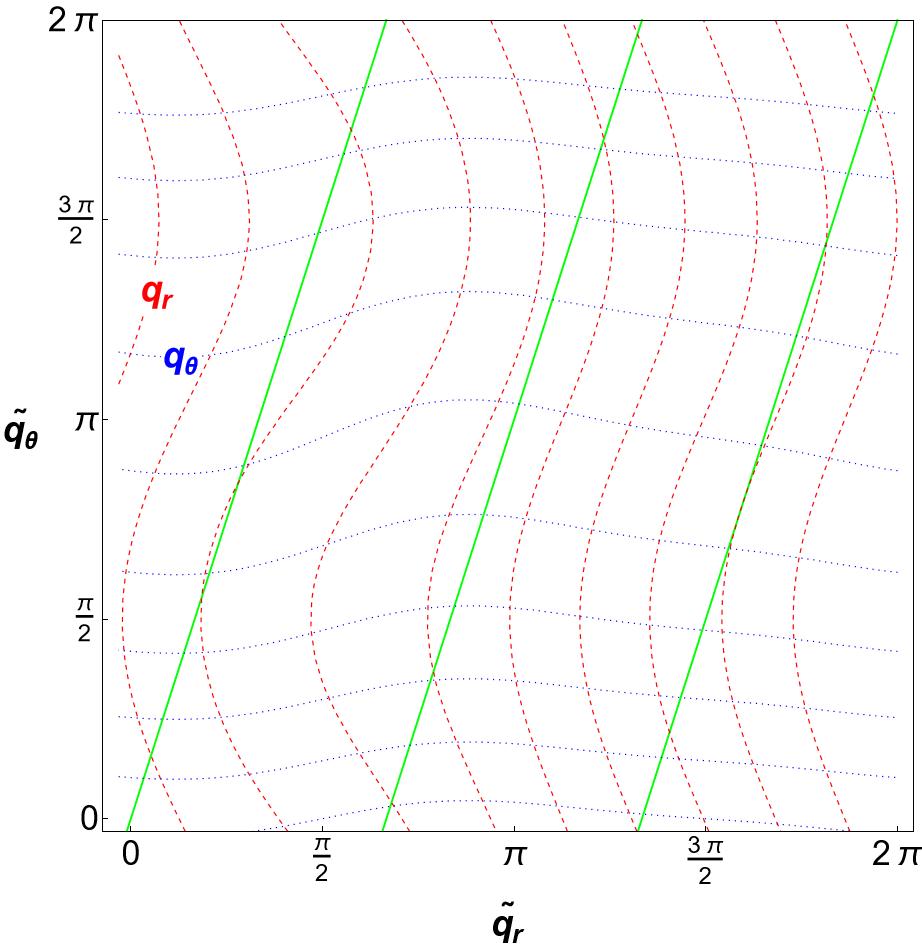}
    \caption{A schematic diagram of the torus and each coordinate. The vertical and horizontal axes represent $\tilde{q}_r$ and $\tilde{q}_\theta$, respectively. The red and blue dotted curves indicate the $q_r=$constant and $q_\theta=$constant contours, and the green solid lines represent the trajectory of the background geodesic orbit. The torus is a geometric object specified by the action variables, which are independent of the choice of the time parameter on ${\cal S}$. Therefore, regardless of which coordinates are used, the rotation numbers of the torus remain unchanged on ${\cal S}$. Consequently, the ratio of frequencies averaged over a long time should be $\langle \Sigma \rangle$.}
    \label{fig:torus1}
\end{figure}

Since the action variables must be identical on the constraint surface ${\cal S}$, we have 
\begin{align}
 \tilde J_a=J_a-\frac12(\mu^2(\bJ)-1)\Delta J_a(\bJ,\tilde \bq)\,.
\end{align}
Here, $\tilde J_a$ is the action variable when $e=\Sigma$. From now on, variables in phase space with Mino time will be denoted with a tilde.
Therefore, we have 
\begin{align}
 \frac{\partial \tilde J_a}{\partial J_b} \approx 
 \delta_a^b+\omega^b\Delta J_a(\bJ,\tilde\bq)\,,
\end{align}
where ``$\approx$'' means that 
the equality holds only on ${\cal S}$. 
The inverse of this Jacobi matrix can be 
calculated as 
\begin{align}
 \frac{\partial J_b}{\partial \tilde J_a} \approx 
 \delta^a_b-\frac{\omega^a\Delta J_b(\bJ,\tilde\bq)}{1+\omega^c\Delta J_c}\,.
\end{align}
Using this expression, we can calculate the 
frequency in Mino time, $\lambda$, as 
\begin{align}
 \tilde \omega^a=\frac{\partial \tilde H}{\partial \tilde J_a} \approx \Sigma 
\frac{\partial J_b}{\partial \tilde J_a}
\frac{\partial H}{\partial J_b}
\approx \frac{\Sigma \omega^a}{1+\omega^c\Delta J_c}\,,
\label{eq:tildeomega}
\end{align}
Both $q^a$ and $\tilde q^a$ are the coordinates on the same torus specified by a given set of $\bJ$ (or equivalently by that of $\tilde \bJ$), and they change by $2\pi$ when a point on the torus moves 
along each cycle $C_a$ \footnote{The $2\pi$ periodicity in time direction is imposed by hand.}.
These frequencies must be related by $\tilde\omega^a=\langle \Sigma\rangle \omega^a$, 
because $2\pi/\omega^a$ and $2\pi/\tilde \omega^a$ are the 
averaged period of orbiting the cycle $C_a$ on the 
torus, respectively, in the proper time and Mino time (see Fig.\ref{fig:torus1}).

Thus, from Eq.\eqref{eq:tildeomega}, we obtain
\begin{align}
 \omega^c\Delta J_c=\frac{{\Sigma}-{\langle\Sigma\rangle}}{\langle\Sigma\rangle}\,. 
 \label{eq1}
\end{align}

Now, we derive the generating function $W$ for 
the canonical transformation between $\{\bJ,\bq\}$ 
and $\{\tilde\bJ,\tilde\bq\}$. 
To satisfy Eq.~\eqref{eq1}, using
\begin{align}
  \tilde J_a=\frac{\partial W(\bJ,\tilde\bq)}{\partial \tilde q^a}\,,
\end{align}
we obtain
\begin{align}
 \omega^a\frac{\partial W(\bJ,\tilde\bq)}{\partial \tilde q^a}
 =\omega^a J_a
 -\frac12(\mu^2-1)\frac{\Sigma-\langle\Sigma\rangle}{\langle\Sigma\rangle}\,.
 \label{eq:Wdef}
\end{align} 
Thus, we find 
\begin{align}
  W=&J_a \tilde q^a -\frac{\mu^2-1}2 f(\bJ,\tilde q^r,\tilde q^\theta)\,.
\label{eq:generatingfucntion}
\end{align}
with
\begin{align}
  &f(\bJ,\tilde q^r,\tilde q^\theta)
  :=\frac1{\tilde\omega^r}\int d\tilde q^r \left[r^2(\bJ,\tilde q^r)-\langle r^2\rangle\right] \cr
 &\quad -
  \frac{a^2}{\tilde\omega^r}\int d\tilde q^\theta \left[\cos^2\theta(\bJ,\tilde q^\theta)-\langle\cos^2\theta\rangle\right]\,.
\end{align}
It is easy to show that the only remaining ambiguity is to add 
a $\tilde\bq$-independent function to $W$, which merely changes the origin of $\tilde\bq$. 
Even if $e(\tau)$ were not separated into the $\tilde q^r$-dependent part and the $\tilde q^\theta$-dependent part,  
we can integrate \eqref{eq:Wdef} to obtain $f$ by expanding 
$e-\langle e \rangle$ into double Fourier series of $\tilde q^r$ and 
$\tilde q^\theta$ as 
\begin{align}
    \frac{e-\langle e\rangle}{\langle e \rangle}
      =\sum_{n^r,n^\theta} \sigma_{n_r n_\theta}\exp(in_r \tilde q^r +in_\theta \tilde q^\theta)\,.
\end{align}
Then, we find $f$ is given by 
\begin{align}
   f(\bJ,\tilde q^r,\tilde q^\theta)
      =\sum_{n^r,n^\theta} \sigma_{n_r n_\theta}\frac{\exp(in_r \tilde q^r +in_\theta \tilde q^\theta)}{i (n_r \tilde\omega^r +n_\theta \tilde\omega^\theta)}\,. \label{eq:f}
\end{align}
It should be noted that the above equation does not account for the resonance where $\omega^r/\omega^\theta$ is rational\cite{Flanagan:2010cd}. Even in the case of resonance, $e$ can be expanded in a Fourier series, but the Fourier components that satisfy the resonance condition become constant. If the summation is taken over non-constant terms only that remain after subtracting $\langle e\rangle$, the right-hand side of \eqref{eq:f} does not diverge. However, discontinuities arise in $f(\bJ,\tilde q^r,\tilde q^\theta)$ as a function of $\bJ$ at resonances, and hence the generating function is no longer differentiable with respect to $\bJ$, preventing the transformation\footnote{Generally, frequencies satisfying resonance conditions are densely distributed over phase space, but in practice, higher-order resonances can be ignored\cite{Lynch:2024ohd}, so only the lower-order resonances need to be considered.}. 
In the case of $e=\Sigma$ 
such a pathological behavior does not occur even at resonances.


From this generating function, we obtain the transformation of 
the angle variables as 
\begin{align}
 q^a=\frac{\partial W}{\partial J^a}
  \approx \tilde q^a +\omega^a f(\bJ,\tilde q^r,\tilde q^\theta)\,.
\end{align}
One can check the consistency as 
\begin{align}
 \frac{dq^a}{d\tau}=\frac1\Sigma\left[ \frac{d\tilde q^a}{d\lambda}
 +\omega^a\left(\Sigma-\langle\Sigma\rangle\right)\right]=\omega^a\,.
\end{align}

When we consider the variation of the averaged action
with respect to the initial phases, 
the variations in two respective schemes are 
related by
\begin{align}
 \delta q_{\rm ini}^a=\delta \tilde q^a_{\rm ini}+\omega^a\delta\tau_{\rm ini}\,.
\end{align}
Namely, the variations are not the same in 
two schemes, but the difference can be absorbed by the shift of the origin of the parameter $\tau$, which does not change the trajectory itself. 
This explains more explicitly why two schemes give the same answer to the averaged change rates of the action variables. 

\section{Quick proof of the correspondence}
There is yet another route to prove the equivalence of the balance formulae in two schemes.
Here, we assume the existence of the mode functions of the metric perturbations 
in the form of the partial wave expansion, $\bh_{\omega lm}$\cite{PhysRevD.10.1070,PhysRevD.11.2042,Toomani:2021jlo}. 
The balance formulae are written in terms of the asymptotic amplitudes of partial waves, $Z_{l\sbn}$\cite{Drasco:2005is,Isoyama:2018sib}, which is defined by 
\begin{align}
 \int H_{\rm int}(\bh_{\omega lm}) ds =\sum_\sbn Z_{l\sbn}\delta(\omega-\omega_{\sbn})\,,
\end{align}
with $\omega_{\sbn}:=n_i\omega^i$ ($i=r,\theta,\phi$) and $n_\phi=m$.
Here, the argument of $H_{\rm int}$ represents 
the substituted metric perturbation.  
Also, it is easy to see  
\begin{align}
 \int \tilde H_{\rm int}(h_{\omega lm}) d\lambda
  &=\int H_{\rm int}(h_{\omega lm}) d\tau\,,\cr
 \delta (\omega-\omega_{\sbn})&=\langle\Sigma\rangle\delta(\tilde \omega-\tilde \omega_{\sbn})\,.
\end{align}
From these relations, we find 
\begin{align}
 Z_{l\sbn}=\langle\Sigma\rangle^{-1}\tilde Z_{l\sbn}\,. 
 \label{eq:Zrelation}
\end{align}
This is an alternative way to show that the formulae obtained in one scheme can be translated to the other. 

It would be with mentioning that the Fourier coefficients 
of $H_{\rm int}(h_{\omega lm})$ defined by 
\begin{align}
 K_{\omega l\sbn}:=\frac1{(2\pi)^4}
   \int d^4\!q\, e^{i\omega q^t -in_i q^i} H_{\rm int}
    (h_{\omega lm})\,,
\end{align}
are related to the coefficients in Mino time. 
As we know 
\begin{align}
  H_{\rm int}(\bh_{\omega lm})&=\Sigma \tilde H_{\rm int}(\bh_{\omega lm})\cr
  & =\Sigma \sum_\sbn d\omega\,  e^{i\omega \tilde q^t -in_i \tilde q^i} \tilde K_{\omega l \sbn}\,,
\end{align}
one can calculate $ K_{\omega l\sbn}$ as 
\begin{align}
 K_{\omega l\sbn}=&
  \frac{\langle\Sigma\rangle^{-1}}{(2\pi)^2}
  \sum_{\sbn'} \int d\tilde q^r d\tilde q^\theta\cr
  &\times 
     e^{-i(n_I-n'_I)\tilde q^I+i(\omega-\omega_\sbn)f(\tilde q^r,\tilde q^\theta)} \tilde K_{\omega l\sbn'}\,,\cr
     \label{eq:Komegaln}
\end{align}
where we use $\det(\partial\bq/\partial \tilde\bq)=\Sigma/\langle\Sigma\rangle$ and $I=r,\theta$. 
Noticing that $Z_{l\sbn}=K_{\omega_{\tbn}l\sbn}/(2\pi)$, we recover the expected relation 
 for the integral along the geodesic~\eqref{eq:Zrelation}, by substituting $\omega=\omega_\sbn$ in Eq.~\eqref{eq:Komegaln}.  

\section{correspondence to the scheme to use the external time as the parameter}
We also often adopt the scheme in which the external time coordinate $t$ is used 
as the parameter along the trajectory of a particle. Let's denote the variables in 
this scheme with hat. This scheme corresponds to choosing $e=(dt/d\tau)^{-1}$. 
The relation between $(\hat \bJ,\hat \bq)$ and $(\tilde \bJ,\tilde \bq)$ is 
completely parallel to the relation between $(\bJ,\bq)$ and $(\tilde \bJ,\tilde \bq)$. 
The key quantity $\Sigma=d\tau/d\lambda$ is now replaced by 
\begin{align}
    \hat\Sigma=\frac{dt}{d\lambda}\approx V_{tr}(\bJ,\tilde q^r)+V_{t\theta}(\bJ,\tilde q^\theta)\,,
\end{align}
with 
\begin{align}
V_{tr} =&-\frac{(r^2+a^2)(J_t(r^2+a^2)+aJ_\phi)}{r^2-2Mr+a^2}\,,\cr
V_{t\theta}  = &a(a J_t \sin^2\theta +J_\phi)\,.
\end{align}
As in the case of $\Sigma$, $\hat \Sigma$ is also given by a sum of 
the $\tilde q^r$-dependent part and the $\tilde q^\theta$-dependent part. 
Thus, the generating function corresponding to  
Eq.~\eqref{eq:generatingfucntion} is obtained as 
\begin{align}
  \hat W&=\hat J_a \tilde q^a-\frac{\mu^2-1}2 \cr 
  & \times \Biggl\{\frac1{\tilde\omega^r}\int d\tilde q^r \left[V_{tr}(\hat\bJ,\tilde q^r)-\langle V_{tr}\rangle\right] \cr
  &\quad -\frac{a^2}{\tilde\omega^\theta}\int d\tilde q^\theta \left[V_{t\theta}(\hat\bJ,\tilde q^\theta)-\langle V_{t\theta}\rangle\right]\Biggr\}\,.
\end{align}

This choice of parameterization allows us to interpret $q^t=s$ as the gauge fixing condition~\cite{Dirac:1964}.
Then, the consistency condition of the gauge condition determines the value of $e$ as 
mentioned above. 
The Poisson brackets of the gauge condition and 
the constraint~\eqref{eq:constraint} do not vanish, and therefore they form a pair of second-class constraints. Then, we can substitute this pair of constraints into the action to reduce the dimensions of the phase space from 8 to 6~, as 
\begin{align}
 S& =\int \left[\hat J_\mu \frac{d\hat q^\mu}{ds} + \frac{e}{2} (\mu^2-1)\right]ds\cr
  &\to \int \left[\hat J_i \frac{d\hat q^i}{dt}+\hat J_t(\hat J_j,\hat q^k)\right]dt\,,
\end{align}
where the $\hat J_t=u_t$ term in the last line, 
which is written as a function of the remaining variables by solving the constraint, 
arises from $\hat J_t d\hat q^t/ds$, and $-J_t$ plays the role of Hamiltonian. 
Under this reduction process, the 4-dimensional torus of constants of motion is reduced to the 3-dimensional one, simply by eliminating $\hat q^t$ direction. Therefore, three of the four cycles $\hat C_a$ remain unchanged, and so are the definitions of $\hat J_k$ and $\hat q^k$ with $k=r,\theta,\phi$. 
Accordingly, the transformation between $(\hat\bJ,\hat\bq)$ and $(\tilde\bJ,\tilde\bq)$ is unchanged, with the identification of $\hat q^t=t$ and $\hat J_t=u_t$. 

It would be obvious that the interaction Hamiltonian 
is simply given by $\hat H_{\rm int} = H_{\rm int}/u^t=\tilde H_{\rm int}(dt/d\lambda)^{-1}$, if 
we do not perform the reduction to 6-dimensional phase space. 
The same result can be obtained even if we consider 
the phase space reduction by expanding 
$-J_t$ as the Hamiltonian. 

\section{Summary}
In this study, we demonstrated that action-angle variables for geodesic motion can be constructed independently of the choice of the time parameter 
and clarified the relation between different choices. We also explained how the Hamiltonian dynamics and the action-angle variables are 
defined when we use Mino time, which has been avoided in formal discussions, while commonly used for actual computations. 
We also demonstrated that the action-angle variables in Mino time can be treated in deriving the balance law as in the case of the proper time. 
Furthermore, we clarified the direct relation between the schemes using Mino time and the coordinate time. 
The latter scheme is important for developing the second-order perturbation. 
This result is expected to improve the transperency of future discussions and further advance the study of the two-body problem in general relativity.

\acknowledgements
   We thank Ryuichi Fujita, Soichiro Isoyama, Jack Lewis, Adam Pound, Norichika Sago, Hiroyuki Nakano, Hidetoshi Omiya and Takuya Takahashi for discussions and useful comments. This work is supported by Grant-in-Aid for Scientific Research under Contract Nos. JP23H00110,  JP20K03928, JP24H00963, JP24H01809 and JST SPRING, Grant Number JPMJSP2110(T.K).

\bibliographystyle{plain}
\bibliography{refer}

\end{document}